\documentclass[sigconf,screen=true,bookmarks=false]{acmart}
\AtBeginDocument{%
  }

\usepackage[utf8]{inputenc} %
\usepackage[T1]{fontenc}    %

\usepackage{pifont}
\usepackage{xcolor}         %
\usepackage{booktabs}
\usepackage{hyperref}
\hypersetup{colorlinks,
    linkcolor=red,citecolor=citecolor,urlcolor=blue}
\usepackage{algorithm}
\usepackage{graphicx}
\usepackage{threeparttable}
\usepackage{multirow}
\usepackage{amsmath}
\usepackage{bm}
\usepackage{bbm}
\usepackage{amsfonts}
\usepackage{amsthm}
\usepackage[noend]{algpseudocode}
\usepackage{enumitem} %
\usepackage[subrefformat=parens,labelformat=parens]{subfig}
\usepackage{colortbl}
\usepackage[normalem]{ulem}
\usepackage{soul}
\usepackage{graphicx} %
\usepackage{booktabs}
\usepackage{algpseudocode}
\usepackage{subfig}
\algtext*{EndIf}

\usepackage{wrapfig}
\usepackage{algorithmicx}

\usepackage{xspace}

\algnewcommand{\REQUIRE}{\item[\algorithmicrequire]}
\algnewcommand{\ENSURE}{\item[\algorithmicensure]}

\algnewcommand{\RETURN}{\State \algorithmicreturn }

\algnewcommand{\algorithmicoptional}{\textbf{Optional:}}
\algnewcommand{\OPTIONAL}{\item[\algorithmicoptional]}

\algrenewcommand\algorithmiccomment[1]{\hfill{\footnotesize$\triangleright$ #1}}

\definecolor{citecolor}{RGB}{34,139,34}
\definecolor{mydarkblue}{rgb}{0,0.08,1}
\definecolor{mydarkgreen}{rgb}{0.02,0.6,0.02}
\definecolor{mydarkred}{rgb}{0.8,0.02,0.02}
\definecolor{mydarkorange}{rgb}{0.40,0.2,0.02}
\definecolor{mypurple}{RGB}{111,0,255}
\definecolor{myred}{rgb}{1.0,0.0,0.0}
\definecolor{mygold}{rgb}{0.75,0.6,0.12}
\definecolor{myblue}{rgb}{0,0.2,0.8}
\definecolor{mydarkgray}{rgb}{0.,0.2,0.2}

\definecolor{lightred}{RGB}{255,235,235}
\definecolor{lightgreen}{RGB}{235,255,235}
\definecolor{lightblue}{RGB}{235,235,255}
\definecolor{lightcyan}{RGB}{235,255,255}
\definecolor{lightmagenta}{RGB}{255,235,255}
\definecolor{lightyellow}{RGB}{255,255,235}

\definecolor{qxkcolor}{RGB}{215,235,255}
\definecolor{softmaxcolor}{RGB}{230,235,255}
\definecolor{probxvcolor}{RGB}{255,255,235}

\definecolor{topkcolor}{RGB}{255,235,235}
\definecolor{zecolor}{RGB}{255,255,235}
\definecolor{dynacolor}{RGB}{235,255,255}

\definecolor{reviewcolor}{RGB}{0,0,200}

\theoremstyle{plain}

\theoremstyle{definition}

\newcommand{\squishlist}{
 \begin{list}{$\bullet$}
  { \setlength{\itemsep}{0pt}
     \setlength{\parsep}{3pt}
     \setlength{\topsep}{3pt}
     \setlength{\partopsep}{0pt}
     \setlength{\leftmargin}{1.5em}
     \setlength{\labelwidth}{1em}
     \setlength{\labelsep}{0.5em} } }

\newcommand{\squishend}{
  \end{list}  }

\newcommand{\name}{\texttt{OptoSynthesizer}\xspace}
\newcommand{\nameplace}{\texttt{Apollo}\xspace}
\newcommand{\nameroute}{\texttt{LiDAR}\xspace}

\newcommand{\nameprism}{\texttt{PRISM}\xspace}

\begin{document}

\title{End-to-End Physical Design Automation \\Flow for
Yield-Optimized Inverse-Designed Large-Scale \\Electronic-Photonic Integrated Circuits
}

\author{Hongjian Zhou}
\email{hzhou144@asu.edu}
\affiliation{%
  \institution{Arizona State University}
  \city{Tempe}
  \state{Arizona}
  \country{USA}
}

\author{Haoyu Yang}
\email{haoyuy@nvidia.com}
\affiliation{%
  \institution{NVIDIA Corp.}
  \city{Austin}
  \state{Texas}
  \country{USA}
}

\author{Haoxing Ren}
\email{haoxingr@nvidia.com}
\affiliation{%
  \institution{NVIDIA Corp.}
  \city{Austin}
  \state{Texas}
  \country{USA}
}

\author{Joaquin Matres}
\email{jmatres@gdsfactory.com}
\affiliation{%
  \institution{GDSFactory}
  \city{Mountain View}
  \state{California}
  \country{USA}
}

\author{Jiaqi Gu}
\email{jiaqigu@asu.edu}
\affiliation{%
  \institution{Arizona State University}
  \city{Tempe}
  \state{Arizona}
  \country{USA}
}

\renewcommand{\shortauthors}{Zhou et al.}

\settopmatter{printacmref=false} %
\renewcommand\footnotetextcopyrightpermission[1]{} %
\pagestyle{plain} %
\pagenumbering{gobble} 
\begin{abstract}
As AI systems scale to multi-chiplet and wafer-level architectures, the demand for ultra-high bandwidth and system scalability has outpaced the capabilities of electrical interconnects and computing units. Large-scale heterogeneous electronic-photonic integrated chiplets (EPICs) provide a promising solution, but their practical adoption is limited by the lack of a unified, fabrication-aware physical design automation stack. At the same time, inverse-designed ultra-compact photonic devices offer orders-of-magnitude improvements in spatial and spectral density, yet remain constrained by insufficient design-for-manufacturing support and yield optimization.
In this work, we present \name, an end-to-end physical design automation flow for yield-optimized, inverse-designed EPICs. It integrates three key components across the physical design pipeline: (1) \name-InvDes, a physical-AI-augmented, digital-twin-assisted photonic inverse design and photonics-aware inverse lithography framework; (2) \name-Place, a GPU-accelerated routing-informed EPIC placer for large-scale routability-optimized layout; and (3) \name-Route, a hierarchical curvy-aware waveguide router with global-planning-assisted electrical-optical co-routing. Together, these toolkits form a seamless flow from EPIC netlists to fabrication-ready, yield-robust GDS layouts. We demonstrate how this framework enables compact large-scale photonic tensor cores and high-bandwidth interconnect fabrics for heterogeneous EPIC platforms, providing a practical foundation for manufacturable large-scale EPICs in next-generation AI systems.
\end{abstract}

\maketitle

\section{Introduction}
\label{sec:Introduction}

The rapid growth of AI workloads~\cite{NP_Nature_ahmed}, data-intensive computing~\cite{NP_Light_Zhou}, and high-bandwidth communications~\cite{NP_DATE2020_popstar} is driving demand for scalable and energy-efficient hardware. 
Electronic-photonic integrated circuits (EPICs) are promising platforms, combining mature electronic control with the high bandwidth, low latency, and massive parallelism of photonics. 
At the same time, inverse-designed ultra-compact photonic devices~\cite{NP_minkov2020inverse} offer orders-of-magnitude improvements in spatial and spectral density, making them attractive building blocks for large-scale programmable and application-specific EPIC systems.

However, practical deployment remains limited by the lack of a \emph{unified, fabrication-aware physical design automation stack}. Current PIC design flows are fragmented across device synthesis, layout generation, routing, and post-layout validation~\cite{zhou2025toward}, with limited support for design-for-manufacturing and yield optimization. As a result, devices and circuits optimized in simulation often degrade after fabrication, especially inverse-designed devices with complex geometries that are highly sensitive to fabrication variation.

To address this gap, an effective design flow should incorporate fabrication variations and nonideal physical effects, such as thermal crosstalk, electrical-photonic interaction, and packaging-induced variability, from the outset rather than treating them as post-design corrections, thereby supporting rapid prediction of fabrication yield and realistic system-level performance.

To this end, we propose an end-to-end physical design automation framework for yield-optimized, inverse-designed large-scale EPICs. \ding{202}~At the device level, the framework combines robust inverse design with photonics-aware inverse lithography (ILT) to construct a robust inverse-designed PDK. \ding{203}~At the circuit level, it integrates GPU-accelerated routing-informed placement, hierarchical curvy-aware waveguide routing, and global-planning-assisted metal routing for PIC layout generation. \ding{204}~The flow is also compatible with GDSFactory~\cite{GDSFactory} to support netlist export and post-layout simulation, enabling rapid evaluation and design-space exploration for large-scale EPIC systems in realistic settings.

The rest of this paper is organized as follows. Section~\ref{sec:background} reviews the background, Section~\ref{sec:framework} presents the proposed framework, and Section~\ref{sec:example} demonstrates its capabilities through three examples.

\begin{figure}
    \centering
    \includegraphics[width=\columnwidth]{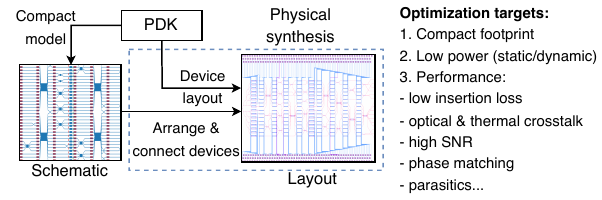}
    \vspace{-15pt}
    \caption{PIC physical synthesis flow: from schematic and PDK-based device layouts to a final connected layout, targeting compact area, low power, and high performance.}
    \label{fig:layoutflow}
    \vspace{-10pt}
\end{figure}

\section{Photonic Physical Synthesis Background}
\label{sec:background}
\subsection{Device-level Inverse Design}
\begin{figure}
    \centering
    \includegraphics[width=0.65\columnwidth]{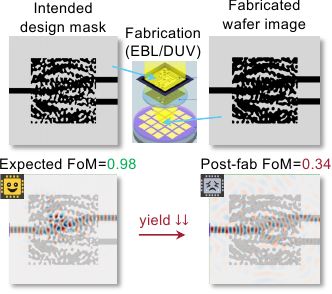}
    \vspace{-5pt}
    \caption{Fabrication error causes severe performance and yield degradation on inverse-designed photonic components.}
    \label{fig:motivation_fab}
    \vspace{-10pt}
\end{figure}

Photonic inverse design formulates device synthesis as an optimization problem that starts from a target figure of merit (FoM), such as transmission and spectral response, and iteratively updates the device geometry using electromagnetic simulation and optimization algorithms, most commonly adjoint-based methods~\cite{NP_khoram2020controlling}. Compared with conventional manual design, which typically tunes a small number of parameters within a predefined topology, inverse design can explore high-dimensional and non-intuitive design spaces, enabling much more compact and higher-performance devices for dense photonic integration. However, despite its strong numerical advantages, inverse-designed photonic devices often suffer from limited manufacturability~\cite{DFM_zhou2026prism}. Their irregular subwavelength geometries are highly sensitive to fabrication variations~\cite{wang2019robust, wang2011robust, schevenels2011robust, gershnabel2022reparameterization}, such as lithography distortion, etch bias, and linewidth variation, which can cause severe performance degradation and low yield after fabrication. Small errors in sensitivity-critical regions can dominate performance even when the overall pixel-wise error is small, as illustrated in Fig.~\ref{fig:motivation_fab}.

 \subsection{Circuit-level Layout Design}
Schematic-driven layout~\cite{NP_book_chrostowski} remains the conventional PIC design flow as shown in Fig~\ref{fig:layoutflow}. Designers first specify circuit functionality and connectivity at the schematic level, and then manually translate the circuit into a physical layout. Unlike electrical interconnects, optical waveguides are highly geometry-sensitive, with direct impact on insertion loss, phase error, and overall circuit performance. As a result, layout generation usually requires careful planning of device placement, waveguide crossings, and path-length/phase matching.
This process typically involves repeated iterations among schematic editing, interactive layout generation, and verification.
While manageable for small circuits, such a manual flow becomes increasingly time-consuming and difficult to scale for large PICs with dense interconnects and complex constraints.

To improve scalability, many works~\cite{NP_DAC2018_planaronoc, NP_DAC2009_ding, NP_ISPD2019_psion} have explored layout automation for PICs, focusing mainly on PIC placement and waveguide routing. These methods aim to reduce manual effort by optimizing congestion, crossings, and insertion loss. However, most existing approaches still focus on abstract planning and cannot directly generate complete real GDS layouts. Their outputs often remain intermediate representations that require substantial manual refinement before tapeout. Bridging the gap between algorithmic optimization and real GDS-level physical realization remains a central challenge in PIC physical synthesis.

\section{Proposed \name Framework}
\label{sec:framework}
\begin{figure}
    \centering
    \includegraphics[width=1\columnwidth]{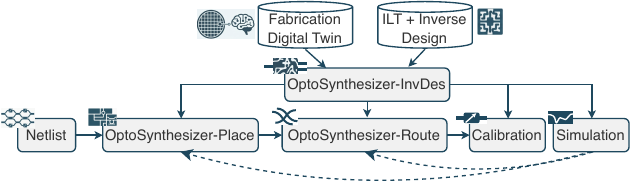}
    \vspace{-10pt}
    \caption{Overview of the proposed fabrication-aware physical design flow \name for inverse-designed PICs.}
    \label{fig:optoflow}
    \vspace{-13pt}
\end{figure}

PIC design has historically relied on expert-driven manual crafting and empirical trial-and-error, leading to slow iterations and limited design-time predictability of post-fabrication behavior, making it unsustainable for large-scale inverse-designed PICs.
We envision a new EPDA paradigm that replaces this workflow with rapid, automation-driven, and yield-native physical synthesis, in which manufacturability and circuit performance are treated as first-class design objectives from device generation to post-layout evaluation.
To realize this vision, we construct a full-stack, end-to-end framework \name (Fig.~\ref{fig:optoflow}) spanning \ding{202}~yield-aware device inverse design with photonic inverse lithography flow for post-fabrication distortion prediction and compensation, \ding{203}~automated PIC schematic-to-layout synthesis with photonics-specific placement and routing engines, and \ding{204}~post-layout and post-fab circuit performance evaluation for yield-aware cross-layer co-design. 
This \name flow establishes a foundation for a standardized and reproducible EPDA methodology for benchmarking and deploying yield-optimized large-scale inverse-designed PIC systems.
We realize this workflow by integrating our EPDA toolboxes~\cite{maps, boson, DFM_zhou2026prism, PLACE_ICCAD2025_Zhou, LiDAR_ISPD_Zhou, LiDAR2_TCAD_Zhou, LiDAR3_ISPD_Zhou} into a unified yield-driven framework for inverse design, automated physical synthesis, and yield-aware evaluation.

\subsection{\name-InvDes}
\label{sec:inv}
\name-InvDes provides foundries and designers with a workflow for creating compact, yield-optimized inverse-designed photonic components. Its central goal is to transform inverse-designed devices from fabrication-fragile geometries into manufacturable, reusable circuit primitives. 
\name-InvDes integrates two complementary capabilities: \uline{fabrication-robust inverse design} for robust-by-construction device synthesis, and \uline{photonics-aware inverse lithography} for physical AI-enabled post-fab prediction and compensation.

\vspace{-5pt}
\subsubsection{Fabrication-Robust Inverse Design}
When creating customized inverse-designed components, we build on MAPS~\cite{maps}, our developed open-source AI-augmented photonic inverse design infrastructure. 
Instead of relying on heuristic minimum-feature-size control, curvature penalties, and simplified over/under-etch modeling, we incorporate differentiable fabrication modeling with variation sampling directly into the end-to-end inverse design loop~\cite{boson}.
This enables efficient optimization within a manufacturable design subspace, where device performance and fabrication robustness are co-optimized under process and operating variations.
The resulting device layout, together with its port information and S-parameter model, is exported and packaged as a PDK cell. 
Our framework adopts a GDSFactory-compatible PDK format~\cite{GDSFactory,Matres2026GDSFactory} to ensure broad interoperability with downstream EDPA ecosystems.

\vspace{-5pt}
\subsubsection{Photonics-aware Inverse Lithography}
As a complementary flow for existing device designs, particularly inverse-designed components, fabrication correction is essential to maintain post-fabrication yield under E-beam lithography (EBL) and, even more critically, under DUV photolithography for high-volume manufacturing.
Fabrication correction is not simply geometric recovery, because small distortions in critical regions can induce large degradation in optical functionality~\cite{DFM_zhou2026prism,DFM_park2026interpretable}.
To address this, we introduce a new concept of \emph{photonics-aware inverse lithography (PA-ILT)} and develop a framework \textit{\nameprism}~\cite{DFM_zhou2026prism}, which learns a physics-augmented AI digital twin to model mask-to-wafer pattern transfer and inversely optimize masks guided by device sensitivity.
The corrected layout can faithfully reproduce the intended behavior after manufacturing, even for complicated structures under 193nm DUV. 
The fabrication digital twin also provides a fast virtual fab surrogate to facilitate post-fab evaluation at the circuit level.

\subsection{\name-Place}
A key challenge in PIC physical design is the strong coupling between placement and routing, since they jointly determine both \emph{PIC performance and routability}. 
Device locations shape interconnect topology and path lengths, affecting insertion loss and crosstalk. 
Meanwhile, photonic routing is constrained by the directional, curvilinear, and area-intensive nature of optical waveguides, which require large bend radii and incur non-negligible crossing overhead. In addition, densely packed and orientation-sensitive photonic ports further limit port accessibility and routing space. These properties make routing-aware placement essential for compact and physically realizable PIC layouts.

In \textit{\nameplace}~\cite{PLACE_ICCAD2025_Zhou}, we address automated PIC placement through \emph{routability optimization, designer-constrained optimization, and scalable acceleration.} To improve routability, Apollo introduces the PIC-specific \textbf{cosWA} wirelength model, which captures port orientation and bend-induced routing overhead, together with a routing-informed \textbf{net spacing model} that preserves whitespace for crossings and dense port access. To satisfy designer intent, \nameplace supports constraint optimization through conditional projection and cell inflation, enabling automatic enforcement of alignment, regularity, and spacing constraints, thereby producing intuitive placement solutions. Built on a GPU-accelerated analytical engine, \nameplace can efficiently optimize large-scale PIC layouts and generate highly routable placements, completing PICs with \textbf{over 1,000 components in about 100 s}. This significantly shortens turnaround time, enables exploration of more layout configurations such as chip area and spacing, and can also provide high-quality initial solutions or guidance for designers.

\vspace{-5pt}
\subsection{\name-Route}
Routing is one of the \textbf{most time-consuming stages} in PIC physical design, and the challenge grows rapidly with circuit scale. As the number of optical and electrical interconnects increases, both waveguide and metal nets must be realized under tight physical constraints. PIC routing is especially resource-limited: optical routing is usually confined to a single silicon waveguide layer, while electrical routing often has access to only a few metal layers. In addition, electrical-optical interaction, such as metal-induced optical loss, and designer-specific spacing rules further complicate routing closure, making PIC routing both resource-constrained and highly application-dependent.

The \nameroute-series PIC routing engine~\cite{LiDAR_ISPD_Zhou, LiDAR2_TCAD_Zhou, LiDAR3_ISPD_Zhou} spans \nameroute 1.0 to \nameroute 3.0, extending routing automation from waveguides to metal wires in active PICs. 
For waveguide routing, \nameroute introduces a curvy-aware hierarchical A$^\star$-based framework that \emph{supports different bend radii and spacing requirements while automatically planning routing resources for crossing insertion and port escape}. By exploiting hierarchy, it also enables routing reuse for repeated modules, accelerating layout generation and producing nearly \textbf{DRV-free real GDS layouts}. In particular, it can complete detailed waveguide routing for a large-scale PIC with \textbf{more than 8,000 nets in only 425\,s}~\cite{LiDAR2_TCAD_Zhou}.
In LiDAR 3.0, we further tackle metal routing in active PICs. \emph{Under limited metal-layer resources, it targets low-via, near-planar routing for millimeter-scale pin-to-I/O connections}. It adopts a multi-stage global planning strategy: escape routing first moves pins and pads into blockage-free regions, and panel-based track assignment then provides guidance for detailed routing, leading to a \textbf{17$\times$ speedup} over conventional routers.

\subsection{Closing the DSTCO Loop: Post-layout/fab Evaluation and Co-Optimization}
This stage closes the device-technology-system co-optimization (DSTCO) loop by linking fabrication-aware device models, physical layout synthesis, and circuit-level performance evaluation.

Existing PIC design flows still \emph{lack integrated support for post-layout extraction/simulation}, and provide even \emph{less support for post-fabrication circuit evaluation}. 
This gap is particularly limiting for inverse-designed and densely integrated PICs, whose performance is highly sensitive to physical layout and fabrication variation.
Meanwhile, conventional flows usually treat layout-induced phase errors as problems to be resolved solely through stricter geometric matching, without \emph{considering active calibration as a must-have degree of freedom for active PIC deployment}.
Our EPDA vision addresses both limitations by \textbf{enabling not only layout- and fabrication-aware circuit evaluation}, but also a more \textbf{realistic design space in which physical layout and active calibration can be jointly considered}.

Specifically, \name converts routed physical layouts into evaluable circuit abstractions by decomposing waveguide paths into standard photonic building blocks, such as bends, straight waveguides, and crossings, and reinserting them into the schematic as a \emph{post-layout netlist}.
These layout-extracted components are then annotated with post-fabrication (with optional ILT correction) S-parameter models predicted by the fabrication digital twin.
Before final circuit simulation, we further introduce a \textbf{post-layout phase-calibration step}, in which active phase shifters are tuned on the extracted fabrication-aware netlist with injected thermal crosstalk to emulate realistic on-chip calibration conditions.
The resulting \emph{fabrication-aware calibrated netlist} is exported in a GDSFactory-compatible representation and simulated using its open-source plugins, such as SAX, to evaluate realistic circuit behavior after layout, fabrication, calibration, thermal crosstalk, and optional ILT correction.
Coupled with our fast place-and-route engine, this capability enables rapid chip-level design space exploration with physically grounded feedback, e.g., density, robustness, and calibration overhead tradeoff analysis.
This closes the loop for device-technology-system co-optimization (DSTCO), allowing device synthesis, fabrication correction, physical layout, and circuit behavior to be optimized cross-layer in a unified workflow.

\section{Case Studies of \name}
\label{sec:example}
\subsection{Example 1: Robust Inverse-Designed Photonic PDK Construction}

\begin{figure}
    \centering
    \includegraphics[width=0.85\columnwidth]{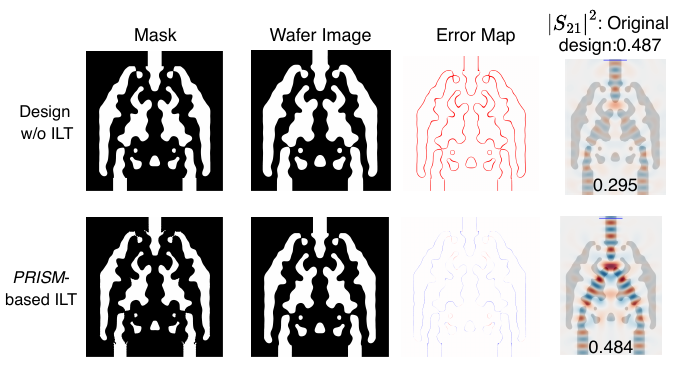}
    \vspace{-10pt}
    \caption{Y-branch comparison with and without PRISM-based inverse lithography under DUV process. ILT reduces mask-to-wafer mismatch and improves $|S_{21}|^2$ from 0.295 to 0.484, close to the design target of 0.487.}
    \label{fig:iltresult}
    \vspace{-10pt}
\end{figure}

These examples are used to demonstrate the capability of our flow. In the first example, we demonstrate the significance of a robust inverse-designed PDK. Using MAPS~\cite{maps}, we first generate compact, high-performance inverse-designed photonic devices. Then, Virtual Fab~\cite{DFM_zhou2026prism} is used to generate post-fabrication data to train a physics-based Hopkins fabrication model~\cite{banerjee2013iccad}, which can better capture the fabrication process and predict wafer patterns from mask layouts.
Table~\ref{tab:duv_ilt} shows that several inverse-designed devices suffer severe performance degradation and near-zero yield under the 193\, nm DUV process. In contrast, \nameprism-based ILT effectively compensates fabrication distortion, substantially recovering device performance and improving yield to near 90\%. Figure~\ref{fig:iltresult} further illustrates a representative Y-branch example, showing that ILT reduces mask-to-wafer mismatch and better preserves the intended performance.

\subsection{Example 2: Thermal-Aware Layout Feasibility Exploration}

\begin{figure}
    \centering
    \includegraphics[width=0.85\columnwidth]{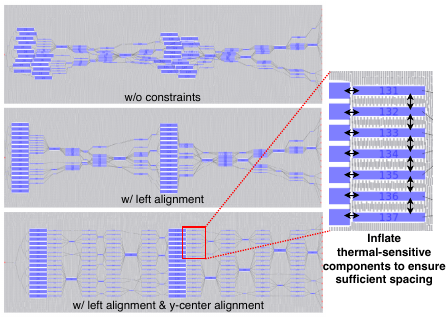}
    \vspace{-10pt}
    \caption{Proper placement constraints ensure high PIC layout quality on ADEPT\_16$\times$16 circuit.}
    \label{fig:placement}
     \vspace{-15pt}
\end{figure}

In this example, we use the proposed flow for early-stage design space exploration by rapidly generating placements and evaluating their performance and physical feasibility. In PIC, thermo-optic phase shifters are highly sensitive to nearby heat sources, making thermal crosstalk a key concern.
Since \nameplace can honor designer-specified constraints and generate placements within minutes, it is well-suited for this task. By adjusting alignment constraints and halo settings, we can control device spacing and layout regularity as shown in Fig.~\ref{fig:placement}. Here, we inflate heater sizes (halo) during placement to enforce extra spacing and study how different halo configurations affect thermal crosstalk and layout feasibility. The thermal crosstalk model follows~\cite{yin2024scatter}. And we further evaluate the output intensity/activation distribution error of the outputs by normalized MSE (NMSE) and the routability~\cite{PLACE_ICCAD2025_Zhou} of the placement result.
Table~\ref{tab:place_opt} shows that, under a fixed design area, increasing the halo of phase shifters reduces thermal crosstalk. 
However, overly small or large halos degrade routability: small halos create local congestion, while large halos consume excessive whitespace and limit routing resources elsewhere. 
The distribution deviation of the output port power decreases as the halo size increases. 
This example shows that our flow enables efficient exploration of spacing configurations and tradeoffs among performance, area, and physical feasibility.

\begin{table}[t]
\caption{Performance and yield improvement from photonic inverse lithography under 193 nm DUV process.}
\vspace{-5pt}
\centering
\resizebox{\columnwidth}{!}{
\begin{tabular}{|c|c|cc|cc|}
\hline
\multirow{2}{*}{Device}          & \multirow{2}{*}{\begin{tabular}[c]{@{}c@{}}Ideal device \\      erformance\end{tabular}} & \multicolumn{2}{c|}{w/o ILT}                & \multicolumn{2}{c|}{PRISM-based ILT}                     \\ \cline{3-6} 
                                 &                                                                                           & Post-fab.            & Yield\_90\%          & Post-fab.            & Yield\_90\%           \\ \hline
Bending                          & $|S_{21}|^2$: 0.964                                                                                 & $|S_{21}|^2$: 0.755±0.052      & 0\%                  & $|S_{21}|^2$: 0.963±0.004      & 100\%                 \\ \hline
Crossing                         & $|S_{31}|^2$: 0.948                                                                                 & $|S_{31}|^2$: 0.718±0.051      & 0\%                  & $|S_{31}|^2$: 0.944±0.005      & 100\%                 \\ \hline
\multirow{2}{*}{Optical diode}   & $|S_{21}|^2$: 0.948                                                                                 & $|S_{21}|^2$: 0.384±0.011      & \multirow{2}{*}{0\%} & $|S_{21}|^2$: 0.886±0.069      & \multirow{2}{*}{85\%} \\
                                 & $|S_{12}|^2$: 0.000                                                                              & $|S_{12}|^2$: 0.122±0.006 &                      & $|S_{12}|^2$: 0.004±0.008 &                       \\ \hline
\multirow{3}{*}{1-to-3 Splitter} & $|S_{21}|^2$: 0.332                                                                                & $|S_{21}|^2$: 0.089±0.026     & \multirow{3}{*}{0\%} & $|S_{21}|^2$: 0.312±0.008     & \multirow{3}{*}{90\%} \\
                                 & $|S_{31}|^2$:   0.326                                                                              & $|S_{31}|^2$: 0.072±0.016     &                      & $|S_{31}|^2$: 0.323±0.020     &                       \\
                                 & $|S_{41}|^2$:   0.323                                                                              & $|S_{41}|^2$: 0.115±0.012     &                      & $|S_{41}|^2$: 0.318±0.014     &                       \\ \hline
\multirow{2}{*}{MDM}             & $|S_{21}|^2$: 0.933                                                                                 & $|S_{21}|^2$:0.696±0.031      & \multirow{2}{*}{0\%} & $|S_{21}|^2$:0.929±0.010      & \multirow{2}{*}{98\%} \\
                                 & $|S_{31}|^2$: 0.940                                                                                 & $|S_{31}|^2$:0.430±0.071      &                      & $|S_{31}|^2$:0.904±0.057      &                       \\ \hline
\end{tabular}
}
\label{tab:duv_ilt}
\vspace{-13pt}
\end{table}

\begin{table}[t]
\caption{Thermal crosstalk and routability of ADEPT\_16$\times$16 under different phase-shifter halo settings.}
\vspace{-5pt}
\centering
\resizebox{8.5cm}{!}{
\begin{tabular}{|cc|c|c|c|c|}
\hline
\multicolumn{2}{|c|}{Phase shifter halo [h\_x, h\_y]}                                                                      & {[}0, 0{]} & {[}35, 15{]} & {[}70, 30{]} & {[}100, 45{]} \\ \hline
\multicolumn{1}{|c|}{\multirow{2}{*}{\begin{tabular}[c]{@{}c@{}}Crosstalk   \\      Coef.\end{tabular}}} & Mean &2.02E-03 & 5.40E-04 & 3.93E-04 & 3.84E-04 \\ \cline{2-6} 
\multicolumn{1}{|c|}{}                                                                                   & Max  & 1.036    & 0.300    & 0.300    & 0.300    \\ \hline
\multicolumn{2}{|c|}{Distribution error (NMSE)}                                                                          & 0.196    & 0.019    & 0.012    & 0.012    \\\hline
\multicolumn{2}{|c|}{Routability}                                                                               & 83.72\%  & 100\%    & 99\%     & 94.98\%             \\ \hline
\end{tabular}
}
\vspace{-5pt}
\label{tab:place_opt}
\end{table}

\subsection{Example 3: Post-Layout PIC Circuit Performance Optimization}
\begin{table}[t]
\caption{Transmission and output distribution deviation (NMSE) of Clements\_16$\times$16 under different correction settings.
}
\vspace{-5pt}
\centering
\resizebox{8.5cm}{!}{
\begin{tabular}{|c|c|c|c|c|}
\hline
Metrics            & Schematic & Post-fab & W/ ILT & W/ ILT \& Phase matching  \\ \hline
Total intensity &0.884    & 0.124    & 0.869    & 0.869      \\ \hline
Distr. deviation   & 0.00 & 9.49E-01 & 1.36E+00 & 3.26E-04    \\ \hline
\end{tabular}
}
\label{tab:phase_matching}
\end{table}

\begin{figure}
    \centering
    \includegraphics[width=0.95\columnwidth]{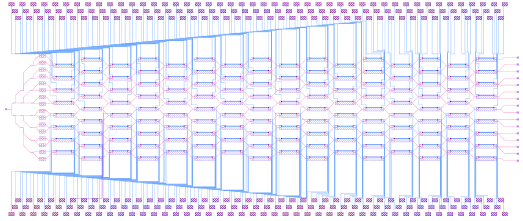}
    \vspace{-10pt}
    \caption{Visualization of \name-generated complete PIC layout of \texttt{Clements\_16$\times$16}.}
    \label{fig:layout_clement}
    \vspace{-5pt}
\end{figure}

In the final example, we use a Clements\_16$\times$16 array~\cite{NP_Optica2018_Clements} to demonstrate rapid netlist-to-layout generation with post-layout and post-fab evaluation. Figure~\ref{fig:layout_clement} shows the layout generated by our flow. 
As a representative coherent computing circuit, the Clements architecture is highly sensitive to fabrication variation and phase mismatch, and thus typically requires post-fabrication calibration. 
With our flow, \textbf{calibration can be considered together with layout optimization}. 

In this experiment, we evaluate the total intensity of the output ports and the intensity distribution error. From Table~\ref{tab:phase_matching}, the post-fabrication circuit shows both severe power degradation and an incorrect output distribution. ILT can recover the transmission power, but the distribution remains incorrect due to phase mismatch. Only with additional phase matching can both be fully restored. In practice, achieving exact phase matching solely through routing is difficult, especially in dense layouts, since even a small phase adjustment may require only a tiny waveguide-length difference (e.g., on the order of 10\, nm), which can be easily disrupted by process variation. Therefore, instead of relying only on geometric matching, our framework enables co-optimization of routing, placement, and post-fabrication calibration to balance layout complexity, calibration overhead, and tuning power, while accelerating layout-aware performance evaluation.

\section{Conclusion}
\label{sec:Conclusion}

This work presents \name, a full-stack, yield-optimized EPDA framework for inverse-designed large-scale PICs.
Moving beyond the traditional manual and fabrication-unaware photonic design paradigm, \name unifies fabrication-robust inverse design, photonic-aware inverse lithography, automated schematic-to-layout physical synthesis, and post-layout/post-fabrication circuit evaluation into a single closed loop. 
Our case studies show that this framework can simultaneously improve device manufacturability, accelerate physical design turnaround, and enable more realistic layout- and calibration-aware circuit optimization. 
More broadly, \name establishes a foundation for standardized, reproducible, and yield-native EPDA, and points toward a new generation of device-technology-system co-optimization workflows for manufacturable large-scale inverse-designed photonic systems.

\clearpage
\balance

\end{document}